\documentclass[letter,bibyear]{aa} 
\usepackage{epsfig}
\usepackage{amsmath}
\usepackage{subfigure}
\usepackage{natbib}
\usepackage{multirow}
\usepackage{color}
\usepackage{longtable}
\usepackage{graphicx}
\usepackage{epstopdf}
\usepackage{booktabs}
\usepackage{txfonts}
\usepackage{graphicx}
\usepackage{txfonts}
\usepackage{color}
\usepackage{stfloats}
\usepackage{amsmath}
\usepackage{multirow}
\usepackage{epsfig}
\usepackage{amsmath}
\usepackage{subfigure}
\usepackage{natbib}
\usepackage{longtable}
\usepackage{graphicx}
\usepackage{epstopdf}
\usepackage{booktabs}
\usepackage{txfonts}
\usepackage[utf8]{inputenc}%
\usepackage{graphicx}
\usepackage{txfonts}
\usepackage{stfloats}

\newcommand{\kms}{km\,s$^{-1}$}

\newcommand{\jms}{J. Mol. Spectr.}
\newcommand{\jmst}{J. Mol. Struct.}

\begin{document}

\title{Identification of the interstellar 1-cyano propargyl radical (HCCCHCN) in TMC-1}

\author{
C.~Cabezas\inst{1} \thanks{Corresponding author: carlos.cabezas@csic.es},
M.~Ag\'undez\inst{1},
N.~Marcelino\inst{2,3},
C.~H.~Chang\inst{4},
R.~Fuentetaja\inst{1},
B.~Tercero\inst{2,3},
M.~Nakajima\inst{5},
Y.~Endo\inst{4},
P.~de~Vicente\inst{2}
and
J.~Cernicharo\inst{1}
}

\institute{Grupo de Astrof\'isica Molecular, Instituto de F\'isica Fundamental (IFF-CSIC), C/ Serrano 121, 28006 Madrid, Spain.\newline \email jose.cernicharo@csic.es
\and Observatorio de Yebes, IGN, Cerro de la Palera s/n, 19141 Yebes, Guadalajara, Spain
\and Observatorio Astron\'omico Nacional (OAN, IGN), C/ Alfonso XII, 3, 28014, Madrid, Spain.
\and Department of Basic Science, Graduate School of Arts \& Sciences, The University of Tokyo, Komaba 3-8-1, Meguro-ku, Tokyo, 153-8902, Japan
\and Department of Applied Chemistry, Science Building II, National Yang Ming Chiao Tung University, Hsinchu 300098, Taiwan
}

\date{Received; accepted}

\abstract
        {We report the first detection in interstellar medium of the 1-cyano propargyl radical, HC$_3$HCN. This species is an isomer of the 3-cyano propargyl radical (CH$_2$C$_3$N), which was recently discovered in TMC-1. The 1-cyano propargyl radical was observed in the cold dark cloud TMC-1 using data from the ongoing QUIJOTE line survey, which is being carried out with the Yebes 40m telescope. A total of seven rotational transitions with multiple hyperfine components were detected in the 31.0–50.4 GHz range. We derived a column density of (2.2$\pm$0.2)$\times$10$^{11}$ cm$^{-2}$ and a rotational temperature of 7$\pm$1\,K. The abundance ratio between HC$_3$HCN and CH$_2$C$_3$N is 1.4. The almost equal abundance of these isomers indicates that the two species may be produced in the same reaction with a similar efficiency, probably in the reaction C + CH$_2$CHCN and perhaps also in the reaction C$_2$ + CH$_3$CN and the dissociative recombination with electrons of CH$_2$C$_3$NH$^+$.}


\titlerunning{Interstellar discovery of HC$_3$HCN}
\authorrunning{Cabezas et al.}

\maketitle

\section{Introduction}

The cold dark cloud TMC-1 has a rich and complex chemistry that leads to the formation of a great variety of molecules. Two molecular line surveys of this cloud, QUIJOTE\footnote{Q-band Ultrasensitive Inspection Journey to the Obscure TMC-1 Environment} \citep{Cernicharo2021a} and GOTHAM\footnote{GBT Observations of TMC-1: Hunting Aromatic Molecules} \citep{McGuire2018}, have detected more than 70 molecular species in the last four years. Recent discoveries include pure hydrocarbons \citep{Cernicharo2021b}, sulphur-bearing species such as HC$_3$S and NC$_3$S \citep{Cernicharo2024a}, and cyano-substituted polycyclic aromatic hydrocarbons such as cyanonaphthalene \citep{McGuire2021}, cyanoacenapthylene \citep{Cernicharo2024b}, and cyanopyrene \citep{Wenzel2024a,Wenzel2024b}. Around 75\% of the latest molecular discoveries in TMC-1 are closed-shell species and the remaining 25\% are open-shell radicals.

Among the radicals detected with QUIJOTE, we can find hydrocarbons as CH$_2$CCH \citep{Agundez2021} and $c$-C$_5$H \citep{Cabezas2022a}, sulphur-bearing radicals such as the mentioned HC$_3$S and NC$_3$S \citep{Cernicharo2024a}, NCS \citep{Cernicharo2021c}, and HCCS$^+$\citep{Cabezas2022b}, and also nitrile-bearing radicals like HC$_3$N$^+$ \citep{Cabezas2024}, HC$_5$N$^+$, and HC$_7$N$^+$ \citep{Cernicharo2024c}, CH$_2$CCN \citep{Cabezas2023}, and CH$_2$CCCN \citep{Cabezas2021a}. Due to their high reactivity, all the cations and radicals have low abundances. However, a special case to note is CH$_2$CCH, which is extremely abundant due to its well-known low reactivity (see e.g. \citealt{Agundez2013}). This added to the spectral dilution resulting from line splitting due to the different interactions between angular momenta makes the detection of radicals fairly challenging. The complexity of the rotational spectrum of a radical is related to the number of nuclei with non-zero nuclear spin. As the number of nuclei with non-zero nuclear spin increases, the prediction and analysis of the fine and hyperfine structure of the rotational spectrum becomes more complicated. Hence, the radio-astronomical discovery of these species is highly dependent on the availability of precise rotational spectroscopic laboratory data. However, thanks to the high sensitivity and resolution of QUIJOTE, some radicals, which display almost trivial spectral patterns, have been identified in TMC-1 without any laboratory data. This is the case for HCCS$^+$\citep{Cabezas2022b}, which has one non-zero nuclear spin nucleus, and  HC$_3$N$^+$ \citep{Cabezas2024}, HC$_5$N$^+$, and HC$_7$N$^+$ \citep{Cernicharo2024b}, which have two non-zero nuclear spin nuclei.

In the case of open-shell radicals with three non-zero nuclear spin nuclei, even very precise ab initio calculations are not enough to unravel their spectral features in the line surveys. Three nuclear spins cause much more complex rotational spectra, with rotational transitions split into dozens of hyperfine components due to the interaction between the rotational, electron spin, and nuclear spin angular momenta. In such cases, the laboratory experiments are invaluable tools, as shown in the identification of the radicals CH$_2$CCN \citep{Cabezas2023} and CH$_2$C$_3$N \citep{Cabezas2021a}. Only aminomethylidyne, H$_2$NC \citep{Cabezas2021b}, a radical with three non-zero nuclear spin nuclei, has been discovered in space without laboratory experimental support.

\begin{figure}
\centering
\includegraphics[angle=0,width=0.3\textwidth]{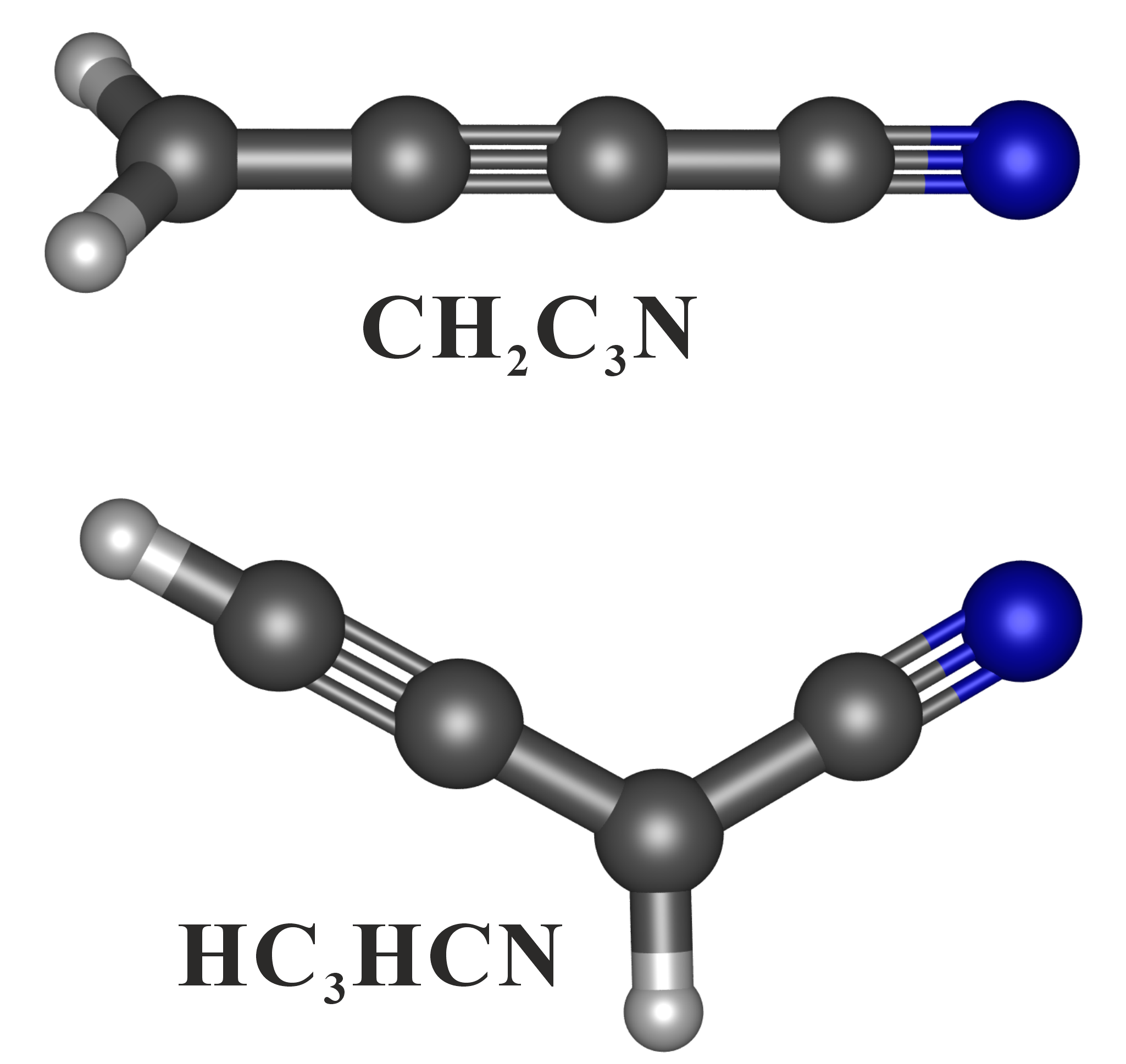}
\caption{Molecular structures of the CH$_2$C$_3$N and HC$_3$HCN radicals.} \label{isomers}
\end{figure}

This work is another example of the astronomical discovery of a radical with a complex rotational spectrum using high-resolution rotational spectroscopy experiment data. The 3-cyano propargyl radical, CH$_2$C$_3$N, was detected in TMC-1 in 2021 by \citet{Cabezas2021a} using the molecular constants derived from laboratory experiments \citep{Chen1998, Tang2001}. One of the routes to form CH$_2$C$_3$N in space is the reaction C + CH$_2$CHCN $\rightarrow$ CH$_2$C$_3$N + H. This reaction has been studied using crossed molecular beam experiments and theoretical calculations \citep{Su2005,Guo2006a}. These studies indicate that the reaction is barrier-less and occurs through H atom elimination, yielding as main products the radicals 3-cyano propargyl (CH$_2$C$_3$N) and also its isomer 1-cyano propargyl (HC$_3$HCN; see Fig. \ref{isomers}). The latter is inferred to be produced at least twice more efficiently than the former \citep{Guo2006a}. On this basis, in 2022 we investigated the rotational spectrum of HC$_3$HCN and carried out an astronomical search of TMC-1 using the QUIJOTE survey \citep{Cabezas2022c}. We did not detect this species at that time but did derive a 3$\sigma$ upper limit to its column density of 6.0$\times$10$^{11}$ cm$^{-2}$, which implies an abundance ratio HC$_3$HCN/CH$_2$C$_3$N\,$<$\,3.8. The non-observation of HC$_3$HCN was rationalized by the large partition function for this radical compared to that for CH$_2$C$_3$N. For example, the partition function at 10\,K for HC$_3$HCN is 9462, while that for CH$_2$C$_3$N is 1718. Hence, the expected intensity of the lines of HC$_3$HCN, assuming the same column density as for CH$_2$C$_3$N, will be much lower, which makes its detection more difficult. However, as stated in our previous work, a deeper integration in TMC-1 would lead to the detection of the HC$_3$HCN radical in this source. The present QUIJOTE dataset provides a more sensitive spectrum than what was used in \cite{Cabezas2022c}, and thus now we revisit the astronomical search for 1-cyano propargyl, reporting the first detection of this radical in space.

\section{Observations}

The astronomical observations presented in this work are from the ongoing Yebes 40m Q-band line survey of TMC-1, the QUIJOTE line survey. A detailed description of the line survey and the data-analysis procedure is provided in \citet{Cernicharo2021a, Cernicharo2022}. Briefly, QUIJOTE consists of a line survey in the Q band (31.0–50.3 GHz) at the position of the cyanopolyyne peak of TMC-1 ($\alpha_{J2000}=4^{\rm h} 41^{\rm  m} 41.9^{\rm s}$ and $\delta_{J2000}=+25^\circ 41' 27.0''$). This survey is being carried out using a receiver built within the Nanocosmos project\footnote{\texttt{https://nanocosmos.iff.csic.es/}} consisting of two cooled high-electron-mobility-transistor amplifiers covering the Q band with horizontal and vertical polarization. Fast Fourier transform spectrometers with $8\times2.5$ GHz and a spectral resolution of 38.15 kHz provide the whole coverage of the Q band in both polarizations. Receiver temperatures are 16\,K at 32 GHz and 30\,K at 50 GHz. The experimental setup is described in detail by \citet{Tercero2021}.

All observations were performed using frequency-switching observing mode with a frequency throw of 10 and 8 MHz. The total observing time on the source for data taken with frequency throws of 10 MHz and 8 MHz is 772.6 and 736.6 hours, respectively. Hence, the total observing time on source is 1509.2 hours. The QUIJOTE sensitivity varies between 0.08 mK at 32 GHz and 0.2 mK at 49.5 GHz. The main beam efficiency can be given across the Q band as $B_{\rm eff}$=0.797 exp[$-$($\nu$(GHz)/71.1)$^2$]. The forward telescope efficiency is 0.97. The telescope beam size at half power intensity is 54.4$''$ at 32.4 GHz and 36.4$''$ at 48.4 GHz. The absolute calibration uncertainty is 10\,$\%$. The data were analysed with the GILDAS package\footnote{\texttt{http://www.iram.fr/IRAMFR/GILDAS}}.

\section{Results}

\begin{figure}
\centering
\includegraphics[angle=0,width=0.47\textwidth]{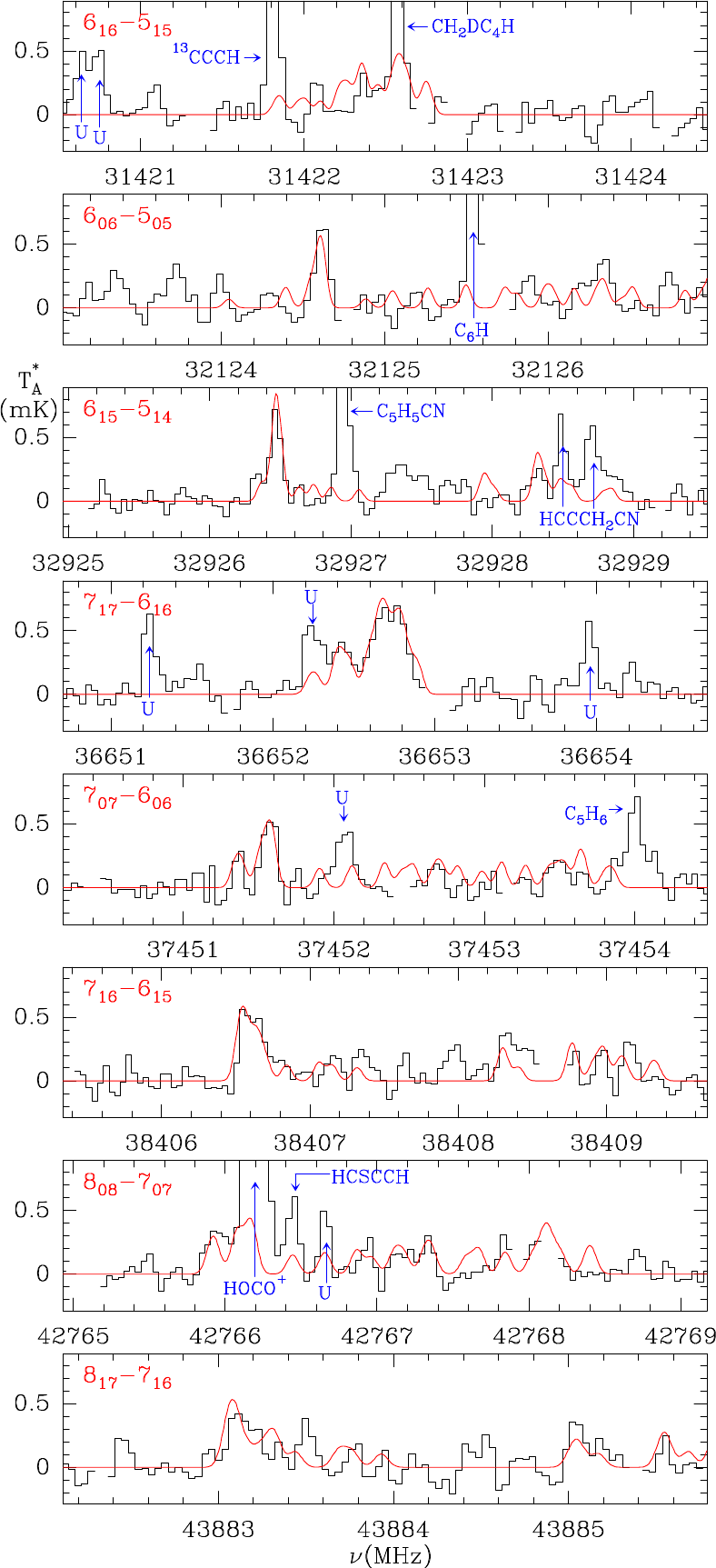}
\caption{Observed lines of HC$_3$HCN in TMC-1 from the QUIJOTE line survey (black histogram) and synthetic spectra (red curve) calculated adopting a column density of 2.2$\times$10$^{11}$ cm$^{-2}$. Blanked channels correspond to negative
features produced in the folding of the frequency-switching data. For each rotational transition, the most intense hyperfine components are shown. The abscissa corresponds to the rest frequency assuming a local standard of rest velocity of 5.83\,\kms. The ordinate is the antenna temperature in millikelvins.} \label{tmc}
\end{figure}

As mentioned before, the rotational spectrum of HC$_3$HCN ($^2A^{''}$) was measured in the laboratory by \cite{Cabezas2022c}. This transient species was produced by electric discharges, and its rotational spectrum was characterized using a Balle-Flygare narrowband-type Fourier-transform microwave spectrometer operating in the frequency region of 4-40 GHz \citep{Cabezas2022c}. The spectral analysis was supported by high-level ab initio calculations. A total of 193 fine and hyperfine components that originated from 12 rotational transitions, $a$ and $b$ type, were measured for the HC$_3$HCN. The analysis allowed us to accurately determine 22 molecular constants, including rotational and centrifugal distortion constants as well as the fine and hyperfine constants. Electric dipole moment components along the $a$ and $b$ inertial axes have been calculated  to be 3.7 and 1.7 D \citep{Cabezas2022c}.

We used the SPCAT program \citep{Pickett1991} to predict the line frequencies from the spectroscopic parameters determined in \cite{Cabezas2022c}. The predictions, which are also available in the CDMS catalogue \citep[entry number 064527;][]{Muller2005}, have an accuracy better than 10 kHz in the Q band (30–50 GHz) for transitions with $K_a$= 0 and 1. Hence, they should be sufficiently accurate and extensive for searches in cold sources. The frequency predictions were implemented in the MADEX code \citep{Cernicharo2012} to compute synthetic spectra assuming local thermodynamic equilibrium. We used the values of the dipole moment components mentioned above and assumed a systemic velocity for TMC-1 of $V_{\rm LSR}$ = 5.83 km s$^{-1}$ \citep{Cernicharo2020}.

Initially, our search focused on $a$-type transitions since their predicted intensity is 4.7 times stronger than that of $b$-type transitions. A total of 12 $a$-type transitions with $K_a$= 0 and 1 are covered by our QUIJOTE survey. The four $N$-triplets, with $N$= 6, 7, 8, and 9, are centred around 32, 37, 43, and 48 GHz. Of these twelve transitions, we detected seven of them well (shown in Fig.\,\ref{tmc}): 6$_{0,6}$-5$_{0,5}$, 6$_{1,5}$-5$_{1,4}$, 7$_{1,7}$-6$_{1,6}$, 7$_{0,7}$-6$_{0,6}$, 7$_{1,6}$-6$_{1,5}$, 8$_{0,8}$-7$_{0,7}$, and 8$_{1,7}$-7$_{1,6}$. For all these transitions, the most intense hyperfine features, resulting from the collapse of several components, are clearly detected, while not all the weak components are well detected. As can be seen, the contamination with other species and with negative features produced in the folding of the frequency-switching data precludes the detection of the weak components. In the case of the 6$_{1,6}$-5$_{1,5}$ transition (top panel of Fig.\,\ref{tmc}), the most intense component overlaps with a strong line of CH$_2$DC$_4$H, but the overall spectral pattern predicted is fully consistent with the observed spectrum. The remaining transitions covered in the Q band, 8$_{1,8}$-7$_{1,7}$, 9$_{1,9}$-8$_{1,8}$, 9$_{0,9}$-8$_{0,8}$, and 9$_{1,8}$-8$_{1,7}$ around 43 and 48 GHz, are predicted with similar intensities to those of the analogue lower-$N$ lines, but they are not detected because the noise is higher in that region of the spectrum due to the increase in the atmospheric attenuation and emissivity by the O$_2$ molecular band around 60 GHz. The $b$-type transitions are expected to be 4.7 times weaker than the $a$-type ones. Hence, their expected emission is below our detection limit. An analysis of the observed intensities using a line profile fitting method \citep{Cernicharo2021d} provides a rotational temperature of 7 $\pm$ 1 K, while the observed intensities are reproduced with a column density of (2.2$\pm$0.2)$\times$10$^{11}$ cm$^{-2}$. The derived column density for HC$_3$HCN is about three times lower than the upper limit estimated in \cite{Cabezas2022c}. Given the column density of CH$_2$C$_3$N in TMC-1, 1.6\,$\times$\,10$^{11}$ cm$^{-2}$ \citep{Cabezas2021a}, the column density ratio between the two isomers, HC$_3$HCN/CH$_2$C$_3$N, is 1.4.

\section{Chemical modelling}

The chemistry of the radical 3-cyano propargyl (CH$_2$C$_3$N) in cold dense clouds has been discussed in \cite{Cabezas2021a}. Here we revisit the chemistry of cyano propargyl radicals in the light of the discovery of the isomer 1-cyano propargyl (HCCCHCN) in \mbox{TMC-1}. To that purpose, we ran chemical modelling calculations, adopting typical parameters of a cold dense cloud: a gas temperature of 10 K, a volume density of H nuclei of 2\,$\times$\,10$^{4}$ cm$^{-3}$, a cosmic-ray ionization rate of 1.3\,$\times$\,10$^{-17}$ s$^{-1}$, a visual extinction of 30 mag, and low metal elemental abundances \citep{Agundez2013}. We used a chemical network from a more recent version of the UMIST database \citep{Millar2024} than that used in \cite{Cabezas2021a}. Unlike in the previous version, the radical CH$_2$C$_3$N is now included in the UMIST network with several of the reactions of formation and destruction discussed in \cite{Cabezas2021a}. However, other isomers such as HCCCHCN are not included, and, since in many of the reactions involving CH$_2$C$_3$N it is difficult to distinguish between different isomers, we considered that CH$_2$C$_3$N comprises all possible C$_4$H$_2$N isomers.

\begin{figure}
\centering
\includegraphics[angle=0,width=\columnwidth]{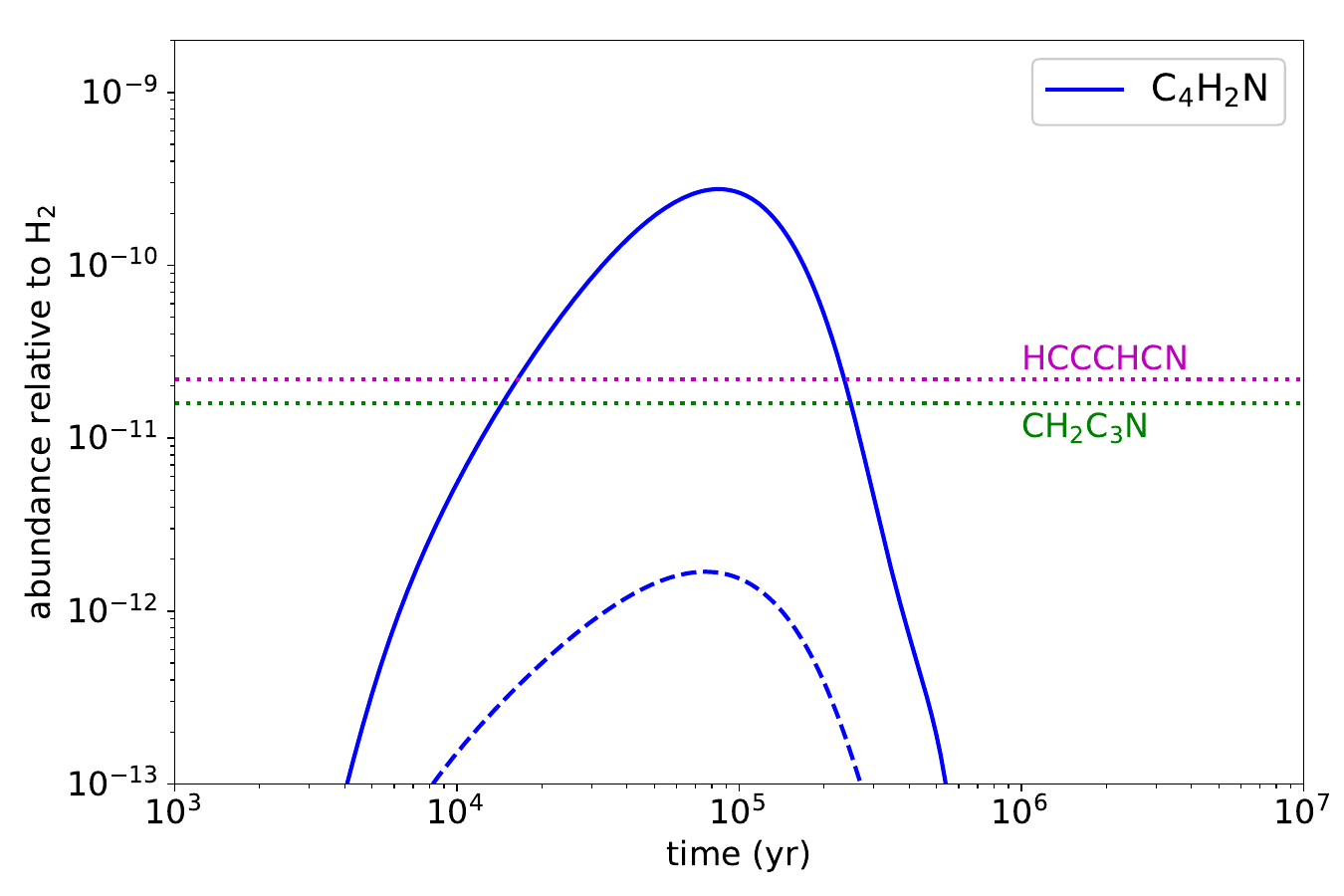}
\caption{Calculated fractional abundance of cyano propargyl radicals (C$_4$H$_2$N, where all possible isomers are included) as a function of time. The solid blue line corresponds to the standard UMIST 2022 case, in which reactions of C$_4$H$_2$N with neutral atoms are not included, while the dashed blue line corresponds to the case in which C$_4$H$_2$N radicals react fast with neutral C, N, and O atoms. The horizontal dotted lines correspond to the observed abundances of the two cyano propargyl isomers detected in \mbox{TMC-1}, CH$_2$C$_3$N, and HCCCHCN, adopting a column density of H$_2$ of 10$^{22}$ cm$^{-2}$ \citep{Cernicharo1987}.} \label{fig:abun}
\end{figure}

The calculated abundance of C$_4$H$_2$N is shown as a function of time in Fig.\,\ref{fig:abun} (solid blue line). It is seen that the peak abundance, reached at a time of around 10$^5$ yr, is a few 10$^{-10}$ relative to H$_2$, while the observed abundances of the two isomers CH$_2$C$_3$N and HCCCHCN lie around one order of magnitude below. That is, the chemical model is too efficient at producing cyano propargyl radicals. This is in contrast with the results obtained in \cite{Cabezas2021a}, where the calculated abundance of CH$_2$C$_3$N remained below the observed value (in the case in which the reaction N + C$_4$H$_3$ was neglected). The reason is that the UMIST 2022 network does not include reactions of destruction of CH$_2$C$_3$N with neutral atoms, such as C, N, and O, which were included in \cite{Cabezas2021a}. If these reactions are included, then the calculated peak abundance of C$_4$H$_2$N decreases by two orders of magnitude (see the dashed blue line in Fig.\,\ref{fig:abun}), in agreement with the calculations shown in \cite{Cabezas2021a}. This illustrates how critical the reactions with neutral atoms are for establishing how abundant certain molecules can be in cold dense clouds. Unfortunately, in the case of cyano propargyl radicals, the kinetics of the reactions with neutral atoms is not known.

According to the chemical model, the main reactions of formation of cyano propargyl radicals (C$_4$H$_2$N) are the neutral-neutral reactions
\begin{equation}
\rm C_2 + CH_3CN \rightarrow C_4H_2N + H, \label{reac:c2+ch3cn}
\end{equation}
\begin{equation}
\rm C + CH_2CHCN \rightarrow C_4H_2N + H, \label{reac:c+ch2chcn}
\end{equation}
and the dissociative recombination of the cation CH$_2$C$_3$NH$^+$ with electrons, which is assumed to yield C$_4$H$_2$N isomers and H atoms as fragments. As far as we know, reaction\,(\ref{reac:c2+ch3cn}) has not been studied theoretically or experimentally, although related reactions have been studied and some information can be gained from such studies. The reaction C + CH$_3$CN has been experimentally found to be rapid at low temperatures, with a rate coefficient in the range (3-4)\,$\times$\,10$^{-10}$ cm$^3$ s$^{-1}$ and with H atom elimination being the main channel, with a measured branching ratio of 0.63 \citep{Hickson2021}. Similarly, the reaction C$_2$ + CH$_3$CCH has been measured to be rapid at low temperatures, with a rate coefficient in the range (4-5)\,$\times$\,10$^{-10}$ cm$^3$ s$^{-1}$ \citep{Daugey2008}, while theoretical calculations and crossed-beam experiments indicate that the H atom elimination channel is also the preferred one, with an estimated branching ratio of 0.65 \citep{Guo2006b,Mebel2006}. The behaviour of the C + CH$_3$CN and C$_2$ + CH$_3$CCH reactions suggests that reaction\,(\ref{reac:c2+ch3cn}) could behave similarly (i.e. fast at low temperatures and with H atom elimination as the main channel). It would be interesting to study this reaction to have evidence on whether or not this is the case, and if so, to determine the branching ratios of the different C$_4$H$_2$N isomers formed. Reaction\,(\ref{reac:c+ch2chcn}) has been studied through theoretical calculations and using crossed-beam experiments \citep{Su2005,Guo2006a}. According to these studies, the reaction is barrier-less and the main channel is again H atom elimination. Moreover, the isomer 1-cyano propargyl (HCCCHCN) is inferred to be  twice more abundant than the 3-cyano propargyl radical (CH$_2$C$_3$N). This later point is of great interest given that in \mbox{TMC-1} 1-cyano propargyl is observed to be 1.4 times more abundant than 3-cyano propargyl. The agreement between \mbox{TMC-1} observations and the study by \cite{Guo2006a} on the C + CH$_2$CHCN reaction and the preference of HCCCHCN over CH$_2$C$_3$N supports a scenario in which this reaction would be a major source of cyano propargyl radicals in cold dense clouds.

In summary, the observed abundance of cyano propargyl radicals in \mbox{TMC-1} and their observed abundance ratio suggest that the reaction C + CH$_2$CHCN is an important source of these radicals. Still, there are various open questions regarding the chemistry of these species, namely: (1) whether or not cyano propargyl radicals react fast with neutral atoms; (2) whether or not the reaction C$_2$ + CH$_3$CN is rapid at low temperatures and occurs through H atom elimination, and if so, what the product branching ratios are; (3) what branching ratios of the channels lead to cyano propargyl radicals in the dissociation recombination of the ions CH$_2$C$_3$NH$^+$ and CH$_3$C$_3$NH$^+$ with electrons; and (4) whether or not the reaction N + C$_4$H$_3$ could be a source of cyano propargyl radicals (see \citealt{Cabezas2021a} for a discussion on this point).

\section{Conclusions}

We have reported the detection of the 1-cyano propargyl radical (HC$_3$HCN) in the cold dark cloud TMC-1. This discovery was achieved using a combination of high-resolution rotational spectroscopy experiments and ultra-sensitive astronomical observations. A total of seven rotational transitions with several hyperfine components were observed in our Q-band survey of TMC-1. We find that this radical is slightly more abundant than its structural isomer 3-cyano propargyl radical (CH$_2$C$_3$N). The investigation of the chemical routes of these radicals indicates that both can be formed in the neutral-neutral reaction C + CH$_2$CHCN, with uncertain contributions from the reaction C$_2$ + CH$_3$CN and the dissociative recombination with electrons of CH$_2$C$_3$NH$^+$.

\begin{acknowledgements}

This work was based on observations carried out with the Yebes 40m telescope (projects 19A003, 20A014, 20D023, 21A011, 21D005, and 23A024). The 40m radiotelescope at Yebes Observatory is operated by the Spanish Geographic Institute
(IGN, Ministerio de Transportes y Movilidad Sostenible). We acknowledge funding support from Spanish Ministerio de Ciencia e Innovación through grants PID2019-106110GB-I00, PID2019-107115GB-C21, PID2019-106235GB-I00, and PID2022-136525NA-I00. We thank the Consejo Superior de Investigaciones Científicas (CSIC) for funding through project PIE 202250I097. The present study was also supported by Ministry of Science and Technology of Taiwan and CSIC under the MoST-CSIC Mobility Action 2021 (Grants 11-2927-I-A49-502 and OSTW200006). Y. Endo thanks Ministry of Science and Technology of Taiwan through grant MOST108-2113-M-009-25.

\end{acknowledgements}


\begin{thebibliography}{}
\tiny
\bibitem[Ag\'undez \& Wakelam(2013)]{Agundez2013} Ag\'undez, M., \& Wakelam, V. 2013, Chem. Rev., 113, 8710
\bibitem[Ag\'undez et al.(2021)]{Agundez2021} Ag\'undez, M., Cabezas, C., Tercero, B., et al. 2021, \aap, 647, L10 
\bibitem[Cabezas et al.(2021a)]{Cabezas2021a} Cabezas, C., Ag\'undez, M., Marcelino, N., et al. 2021a, A\&A, 654, L9 
\bibitem[Cabezas et al.(2021b)]{Cabezas2021b} Cabezas, C., Ag\'undez, M., Marcelino, N., et al. 2021b, A\&A, 654, A45 
\bibitem[Cabezas et al.(2022a)]{Cabezas2022a} Cabezas, C., Agúndez, M., Fuentetaja, R., et al. 2022a, \aap, 663, L2 
\bibitem[Cabezas et al.(2022b)]{Cabezas2022b} Cabezas, C., Ag\'undez, M., Marcelino, N., et al.\ 2022b, \aap, 657, L4 
\bibitem[Cabezas et al.(2022c)]{Cabezas2022c} Cabezas, C., Nakajima, M., Chang, C. H., et al.\ 2022c, \aap, 657, A24 
\bibitem[Cabezas et al.(2023)]{Cabezas2023} Cabezas, C., Tang, J., Agúndez, M., et al. 2023, \aap, 676, L5 
\bibitem[Cabezas et al.(2024)]{Cabezas2024} Cabezas, C., Agúndez, M., Endo, Y., et al. 2024, \aap, 687, L22 
\bibitem[Chen et al. (1998)]{Chen1998} Chen, W., McCarthy, M. C., Travers, M. J., et al., 1998, ApJ, 492, 849
\bibitem[Cernicharo \& Gu\'elin(1987)]{Cernicharo1987} Cernicharo, J. \& Gu\'elin, M. 1987, 176, 299
\bibitem[Cernicharo(2012)]{Cernicharo2012} Cernicharo, J. 2012, in European Conference on Laboratory Astrophysics, eds. C. Stehl\'e, C. Joblin, \& L. d'Hendecourt, EAS Publication Series, 58, 251
\bibitem[Cernicharo et al.(2020)]{Cernicharo2020} Cernicharo, J., Marcelino, N., Pardo, J. R., et al. 2020, A\&A, 641, L9
\bibitem[Cernicharo et al.(2021a)]{Cernicharo2021a} Cernicharo, J., Ag\'undez, M., Kaiser, R., et al. 2021a, \aap, 652, L9 
\bibitem[Cernicharo et al.(2021b)]{Cernicharo2021b} Cernicharo, J., Ag\'undez, M., Cabezas, C. et al. 2021b, \aap, 649, L15 
\bibitem[Cernicharo et al.(2021c)]{Cernicharo2021c} Cernicharo, J., Cabezas, C., Ag\'undez, M., et al. 2021c, \aap, 648, L3 
\bibitem[Cernicharo et al.(2021d)]{Cernicharo2021d} Cernicharo, J., Cabezas, C., Endo, Y., et al. 2021d, \aap, 646, L3 
\bibitem[Cernicharo et al.(2022)]{Cernicharo2022} Cernicharo, J., Ag\'undez, M., Fuentetaja, R., et al. 2022, \aap, 663, L9 
\bibitem[Cernicharo et al.(2024a)]{Cernicharo2024a} Cernicharo, J., Cabezas, C., Agúndez, M., et al. 2024a, \aap, 688, L13 
\bibitem[Cernicharo et al.(2024b)]{Cernicharo2024b} Cernicharo, J., Cabezas, C., Fuentetaja, R., et al. 2024b, \aap, 690, L13 
\bibitem[Cernicharo et al.(2024c)]{Cernicharo2024c} Cernicharo, J., Cabezas, C., Agúndez, M., et al. 2024c, \aap, 686, L15 
\bibitem[Daugey et al.(2008)]{Daugey2008} Daugey, N., Caubet, P., Bergeat, A., et al. 2008, PCCP, 10, 729
\bibitem[Guo et al.(2006a)]{Guo2006a} Guo, Y., Gu, X., Zhang, F., et al. 2006a, PCCP, 8, 5454
\bibitem[Guo et al. (2006b)]{Guo2006b} Guo, Y., Gu, X., Balucani, N., \& Kaiser, R. I. 2006b, J. Phys. Chem. A, 110, 6245
\bibitem[Hickson et al.(2021)]{Hickson2021} Hickson, K. M., Loison, J.-C., \& Wakelam, V. 2021, ACS Earth Space Chem., 5, 824
\bibitem[McGuire et al.(2018)]{McGuire2018} McGuire, B. A., Burkhardt, A. M., Kalenskii, S., et al. 2018, Science, 359, 202
\bibitem[McGuire et al.(2021)]{McGuire2021} McGuire, B. A., Loomis, R. A., Burkhardt, A. M., et al. 2021, Science, 371, 1265
\bibitem[Mebel et al.(2006)]{Mebel2006} Mebel, A. M., Kislov, V. V., \& Kaiser, R. I. 2006, J. Phys. Chem. A, 110, 2421
\bibitem[Millar et al.(2024)]{Millar2024} Millar, T. J., Walsh, C., Van de Sande, M., \& Markwick, A. J. 2024, \aap, 682, A109
\bibitem[M\"uller et al.(2005)]{Muller2005} M\"uller, H.~S.~P., Schl\"oder, F., Stutzki, J., Winnewisser, G. 2005, \jmst, 742, 215 
\bibitem[Pickett (1991)]{Pickett1991} Pickett, H.M. 1991, \jms, 148, 371
\bibitem[Su et al.(2005)]{Su2005} Su, H.-F., Kaiser, R.I., \& Chang, A.H.H. 2005, \jcp, 122, 074320
\bibitem[Tang et al. (2001)]{Tang2001} Tang, J. Sumiyoshi, Y., \& Endo, Y., 2001, ApJ, 552, 409
\bibitem[Tercero et al.(2021)]{Tercero2021} Tercero, F., L\'opez-P\'erez, J. A., Gallego, J. D., et al. 2021, \aap, 645, A37
\bibitem[Wenzel et al. (2024a)]{Wenzel2024a} Wenzel G., Speak, T. H., Changala, P. B. et al. 2024a, Nat. Astron, DOI: 10.1038/s41550-024-02410-9
\bibitem[Wenzel et al. (2024b)]{Wenzel2024b} Wenzel G., Cooke, I. R., Changala, P. B., et al. 2024b, Science, 386, 810
\end{thebibliography}
\end{document}